\documentclass[12pt,preprint]{aastex}

\usepackage{natbib}
\usepackage{graphicx}
\usepackage{amssymb}

\shortauthors{Yelles Chaouche et al.}
\shorttitle{3D structure of active filament from the photosphere and the chromosphere}

\begin{document}

\title{The 3D structure of an active region filament as extrapolated from photospheric and chromospheric observations }

   \author{L. Yelles Chaouche\altaffilmark{1,2},  C. Kuckein\altaffilmark{1,2}, V.
   Mart\'inez Pillet\altaffilmark{1,2},  F. Moreno-Insertis\altaffilmark{1,2}}

\altaffiltext{1}{Instituto de Astrofisica de Canarias, Via Lactea, s/n,
  38205  La Laguna (Tenerife), Spain}
\altaffiltext{2}{ Dept.~of Astrophysics, Universidad de La
  Laguna, 38206 La Laguna (Tenerife), Spain}

\date{}

\begin{abstract}

The 3D structure of an active region (AR) filament is studied using nonlinear
force-free field (NLFFF) extrapolations based on simultaneous observations at a photospheric 
and a chromospheric height. To that end, we used 
the \ion{Si}{1} 10827 \AA\ line and the \ion{He}{1} 10830 \AA\ triplet obtained with the Tenerife Infrared Polarimeter (TIP) at
the VTT (Tenerife). The two extrapolations have been carried out independently 
from each other and their respective spatial domains overlap in a considerable height range. This opens up new possibilities for diagnostics in addition to the usual ones obtained through a single extrapolation from, typically, a photospheric layer. Among those possibilities, this method allows the determination of 
an average formation height of the \ion{He}{1} 10830 \AA\ signal of $\approx 2$ Mm above
the surface of the sun. It allows, as well, to cross-check the obtained 3D magnetic
structures in view of verifying a possible deviation from the force-free
condition especially at the photosphere.
The extrapolations yield
a filament formed by a twisted flux rope whose axis is located at about $1.4$ Mm
above the solar surface. The twisted field lines make slightly more than one
turn along the filament within our box, which results in 0.055 turns/Mm. The
convex part of the field lines (as seen from the solar surface) constitute dips where 
the plasma can naturally be supported. The obtained 3D magnetic structure of
the filament depends on the choice of the observed horizontal magnetic field
as determined from the $180^{\circ}$ solution of the azimuth. We derive a
method to check for the correctness of the selected $180^{\circ}$ ambiguity
solution.

\end{abstract}

\keywords{Sun: filament --- Sun: activity --- Extrapolations.}

\section{Introduction}\label{sec:introduction}

Active region filaments are observed above polarity inversion lines (PIL) where the vertical component of the magnetic field 
changes sign separating the two opposite polarities \citep{babcock55}. They are called filaments when observed on the solar disc and prominences when observed
above the solar limb, though the two terms refer to the same phenomenon \citep[see] [for a review on filaments and prominences]{demoulin98, mackay2010}. Filaments/prominences are formed by plasma 
which has been lifted up above the solar surface and is cooler and denser than its surroundings.
The necessary force to
sustain this plasma is of magnetic origin. 
Here we shall focus on active region (AR) filaments, which are characterized by a strong horizontal component of the magnetic field along the filament 
axis reaching several hundred gauss \citep{kuckein09, kuckein11, guo2010, canou10, jing10}. These filaments  typically lie low
above the solar surface \citep{lites05} compared to quiescent filaments.

There are two classes of models that aim at describing the magnetic structure of filaments: the sheared magnetic arcade model, and the twisted flux 
rope (TFR) one. The former involves photospheric motions of magnetic footpoints near the PIL produced by various mechanisms such as
photospheric flows, vortical motions or the emergence of the upper part of a flux rope \citep{antiochos94, devore00, aulanier02, welsch05, devore05}. The flows force the magnetic field to get reconfigured creating a variety of field
configurations, as dipped field lines \citep{mackay2010}.

The twisted flux rope model assumes that the filament is formed by a dipped flux rope where the material can naturally get located and lifted 
up \citep{vanballe89, leka96, titov99, amari99, lites05, lites10}.
Flux emergence of a twisted flux rope to the solar atmosphere has also been numerically modelled \citep{magara03, fan04, archontis04, amari04, amari05, galsgaard05, cheung07, sykora08, tortosa09, fan09, mactaggart10}.
From observations, \citet{okamoto08, okamoto09} have shown that the change in the horizontal field from normal to inverse configuration can be interpreted as the signature of an emerging flux rope. This has been shown as well in 3D MHD simulations of an emerging flux rope, where the spectropolarimetric signal exhibits similar changes of the horizontal field from normal to inverse \citep{yelles09a} using the same pair of Iron lines \ion{Fe}{1} 6301 \AA\ and \ion{Fe}{1} 6302 \AA\ as the ones of Hinode  SP/SOT. 
  
The numerical models aiming to simulate the properties of filaments assume a
certain number of conditions, like, e.g.: the mere pre-existence of a 
flux rope in the convection zone or in the atmosphere; the amount of twist necessary for a given flux rope to emerge into the atmosphere; 
its morphology and emergence velocity; etc. These assumptions and conditions can hardly be directly tested using  observed
magnetograms only, and it turns out to be very helpful to use extrapolations in order to get information on the observed 3D magnetic field and its
properties. This gives a feed-back to numerical models and allows fine tuning them.

%

In recent years, extrapolations have
played an important role in probing active region filaments properties \citep{guo2010, canou10, jing10}. This is due to (a) the
progress in mathematical and computing techniques of extrapolations \citep[e.g.][]{amari97, wiegelmann04, wiegelmann06, schrijver06, aly07,metcalf08} and (b) to the large improvement in spectropolarimetric
facilities allowing to measure the full stokes vector with ever higher
sensitivity and resolution (E.g. SP/SOT onboard Hinode, TIP at the VTT, 
THEMIS/MTR, etc) see e.g. \citet{lopez06, okamoto08, kuckein09, lites10},
and the review by \citet{mackay2010} for further references.

In the past, extrapolations of active region filaments have exclusively been computed using photospheric fields as boundary values.
Knowing that the photospheric plasma is not completely force free, it is important to compare these extrapolations with ones of the same region but using
chromospheric magnetic field as boundary values for the extrapolations. The
plasma $\beta$ in the chromosphere is usually much smaller than in the
photosphere, so that the magnetic field in the former must be closer to
force-free than in the latter. Simultaneous observations of the magnetic
field vector at the photosphere and chromosphere \citep{kuckein11, sasso11} offer the possibility of such a study. Based on the observations
in \citet{kuckein11}, \citep[see also][]{kuckein09, kuckein10} we will perform
extrapolations starting from the photosphere and, independently, from the chromosphere.
This allows to cross check the extrapolation results, and test whether the
latter have been affected by e.g. the finite plasma $\beta$ and the
preprocessing procedure.

We will focus here on two representative times (one for each day of observation). 
Then for each of them we have two snapshots: one at the photosphere using the \ion{Si}{1} 10827 \AA\ line and one at the
chromosphere using the \ion{He}{1} 10830 \AA\  triplet.   
The time evolution of the filament over the entire observation period can be further 
retrieved from the paper by \citet{kuckein11}.

\section{Observations and data analysis}\label{sec:data}

In this paper we study a filament located along the polarity inversion line
of active region NOAA 10781, which was found to be in its slow decay phase when
the observations were performed. The full Stokes spectropolarimetric data were
acquired with the Tenerife Infrared Polarimeter \citep[TIP-II;][]{collados07}
at the German Vacuum Tower Telescope (VTT, Tenerife, Spain) on the 3rd and
5th of July 2005, at coordinates N16-E8 and N16W18, respectively. Several scans were taken
with the slit (0\farcs5 wide and 35\arcsec ) parallel to the filament obtaining
maps with a field of view (FOV) of $\sim 26 \times 25$\,Mm and $\sim 22 \times
25$\,Mm, for July 3rd and 5th respectively. It is important to
note that the FOV was not centered on the same location on both days. However, the upper half of the
map from July 3rd overlaps with the lower half of the map from July 5th. The
observed spectral range comprises the photospheric \ion{Si}{1} 10827\,\AA\ line
and the chromospheric \ion{He}{1} 10830\,\AA\ triplet, which allows us to study
simultaneously both heights and their magnetic coupling. The reader is referred
to the papers of \citet{kuckein09, kuckein11} for an extensive description of the
data and the magnetic evolution of this active region filament.

\begin{figure*} 
\begin{center}
\includegraphics[width=1.0\textwidth]{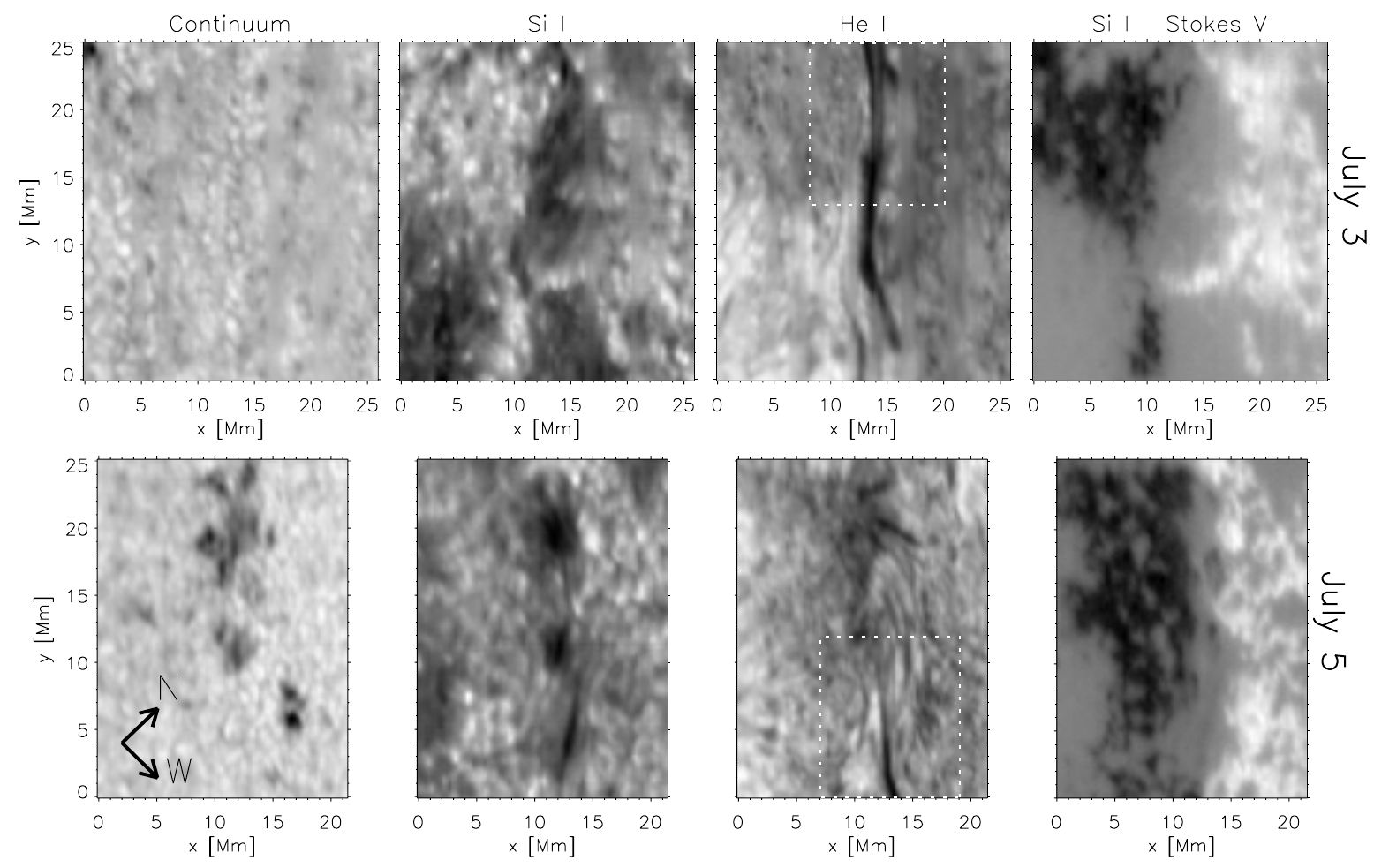}
\end{center}
\caption{Two snapshots: The upper row ($3^{rd}$ of July); The lower row ($5^{th}$ of July). From left to right: Continuum intensity, \ion{Si}{1} 10827 \AA\ line core, \ion{He}{1} 10830 \AA\ red core, and 
Stokes-$V$ at -150 m\AA\ from \ion{Si}{1} 10827 \AA\ line center. As an example, the location of the spine (i.e. the filament's axis) can be seen in the upper He I panel as an elongated darkening along
the y-axis. This can also be seen in the lower He I and \ion{Si}{1} panels. The dashed-line rectangles indicate the approximate overlap of the filament between the two snapshots.} \label{fig1}
\end{figure*}

Figure~\ref{fig1} shows slit-reconstructed maps of the two selected data sets. From left to
right different wavelengths are presented: continuum, Silicon line core
intensity, Helium red core intensity and Silicon Stokes-$V$ magnetogram (at
$-150$\,m\AA\ from line center). The upper map was taken on the 3rd of July 2005,
between 14:39--15:01, and clearly shows the filament in the \ion{He}{1}
absorption panel (third panel starting from top left). The filament extends
along the vertical axis and lies on top of the polarity inversion line, as shown
in the upper Stokes-$V$ panel. The lower map was observed on the 5th of July 2005,
between 8:42--9:01. The spine or filament axis is still seen in the lower half
of the \ion{He}{1} panel, but above, a more diffuse filament is located on top of the
pores and orphan penumbrae which have newly formed. 

Two different inversion codes were used to fit the Stokes profiles. A binning
in the spectral and spatial domain was applied to increase the
signal-to-noise ratio. Hence, the final spectral sampling is $\sim
33.1$\,m\AA\,px$^{-1}$ and the average pixel size is $\sim
1\arcsec$. For the Helium 10830\,\AA\ triplet we used a Milne-Eddington
inversion code \citep[MELANIE;][]{socas01} and for the Silicon 10827\,\AA\ line we used the SIR code
\citep[Stokes Inversion based on Response functions;][]{ruiz92}. \citet{kuckein11} showed that the non local thermodynamical equilibrium (NLTE) effects of the
\ion{Si}{1} 10827\,\AA\ line reported by \citet{bard08} do not
significantly affect the inferred vector magnetic field from the SIR inversion
code and therefore were not taken into account. We transformed the vector
magnetic field from the line of sight (LOS) into the local solar reference
frame. The 180$^\circ$ ambiguity was
solved using the AZAM code \citep{lites95}. Response functions to magnetic
field perturbations for the inferred atmospheres were computed at different
positions inside and around the filament. An average optical depth of $\log \tau
\sim -2$ for the formation height of Silicon has been obtained for both days.

\section{Force-free field Extrapolation}

Using the photospheric or chromospheric magnetic field as boundary values, it is possible to
calculate the 3D magnetic field vector in the atmosphere using force-free field 
extrapolations (FFF). The force-free (or zero-Lorentz force) assumption is
equivalent to assuming that the electrical current $\mathbf{j}$ and the magnetic field
$\mathbf{B}$ are parallel, or, using Amp\`ere's law and cgs units:
\begin{eqnarray}
  \nabla\times \mathbf{B} = \frac{4\,\pi}{c}\mathbf{j} = \alpha\, \mathbf{B} \label{eqn:exp1}\;;
\end{eqnarray} 
$\alpha$ in Eq.~\ref{eqn:exp1} is the so-called force-free parameter and
measures the level of twist of the field lines. Given that $\mathbf{B}$ is
solenoidal, $\alpha$ must be constant along each field line. 
We use here nonlinear force-free extrapolations, where $\alpha$ is constant along
each given field line but can change from one field line to another. 
The extrapolations are carried out using the optimization method \citep{wiegelmann04}.  

The observed magnetic field is not necessarily force free. This is especially the case in the photosphere where
the plasma $\beta$ is not small (with $\beta$ the ratio of gas
to magnetic pressures). For instance $\beta \approx 1$ in the lower part of the photosphere \citep[][from calculations using 3D MHD simulations]{yelles09b}.  

\citet{schrijver06, schrijver08, metcalf08, derosa09} have implemented a variety of tests to compute the coronal magnetic field using several NLFFF extrapolation codes. The different codes were tested using analytical, numerical, and observational models. Among other conclusions, these papers have addressed the issue that the NLFFF extrapolations are sensitive to the boundary conditions (usually, photospheric vector magnetograms). These might suffer from uncertainties due to low signal-to-noise conditions, inaccuracies in the resolution of the $180^{\circ}$ ambiguity, or more importantly from non-force-free conditions of the photospheric plasma. 

In order to make the observed magnetic field maps more consistent with the force-free 
assumption and greatly enhance the correctness of the force-free extrapolations it is necessary to apply some preprocessing \citep{wiegelmann06, metcalf08}.
This consists of minimizing the magnetic force and torque \citep{aly89}, minimizing the difference 
between the observed magnetic field and the preprocessed one, and applying a smoothing operator
to remove the small scale variations of the magnetic field.
The preprocessing routine used here is the one developed by \cite{wiegelmann06}.
It is useful to quantify the change produced on the magnetograms by the preprocessing procedure. 
For that we use the usual vector correlation \citep{schrijver06} between the observed vector magnetic field and the preprocessed one. 
It is found that for the case of the \ion{Si}{1} 10827 \AA\  vector magnetogram the correlation between 
the original and preprocessed field is approximately $0.84$. In the case of the \ion{He}{1} 10830\,\AA\ vector magnetogram, the correlation 
is about $0.90$ between the observed and preprocessed magnetic field. This indicates that the observed \ion{He}{1} 10830\,\AA\ magnetic field is closer to a force-free condition than the \ion{Si}{1} 10827 \AA\ case.

The observed vector magnetograms have a relatively small field-of-view and are not isolated from the rest of the active region. 
This might introduce some errors in the NLFFF extrapolated model. Some authors proceed in enlarging the field-of-view by embedding the observed vector magnetograms in MDI maps, nevertheless this might introduce inconsistencies since the MDI magnetograms contain only the line-of-sight component of the magnetic field, and therefore would inconsistently connect magnetic field lines throughout the NLFFF model. It is therefore safer to just use the existing vector magnetograms (after preprocessing) as lower boundary for the NLFFF extrapolations. A further test of the effect of reduced field-of-view on the extrapolation results is carried out in Sec.~\ref{sec:46}. 

The extrapolation starts with potential lateral and top boundaries, whereas the bottom one is the preprocessed observed vector magnetogram. A description of the code can be found in \citep{wiegelmann04, schrijver06, metcalf08}. The NLFFF extrapolated model is expected to be more correct in the lower central part of the computational domain \citep{schrijver06}.

The extrapolations are computed in a box of
$33$ $\times$ $38$ $\times$ $38$ grid points in the x, y and z directions respectively. x and y being the
horizontal directions perpendicular and parallel to the filament's axis respectively. This yields a distance of $\approx 650$ km between
grid points. The same distance between grid points is used in the
vertical direction. 

\section{Results}
\subsection{Extrapolations starting from the photosphere}

\begin{figure*} 
\begin{center}
\includegraphics[width=1.0\columnwidth]{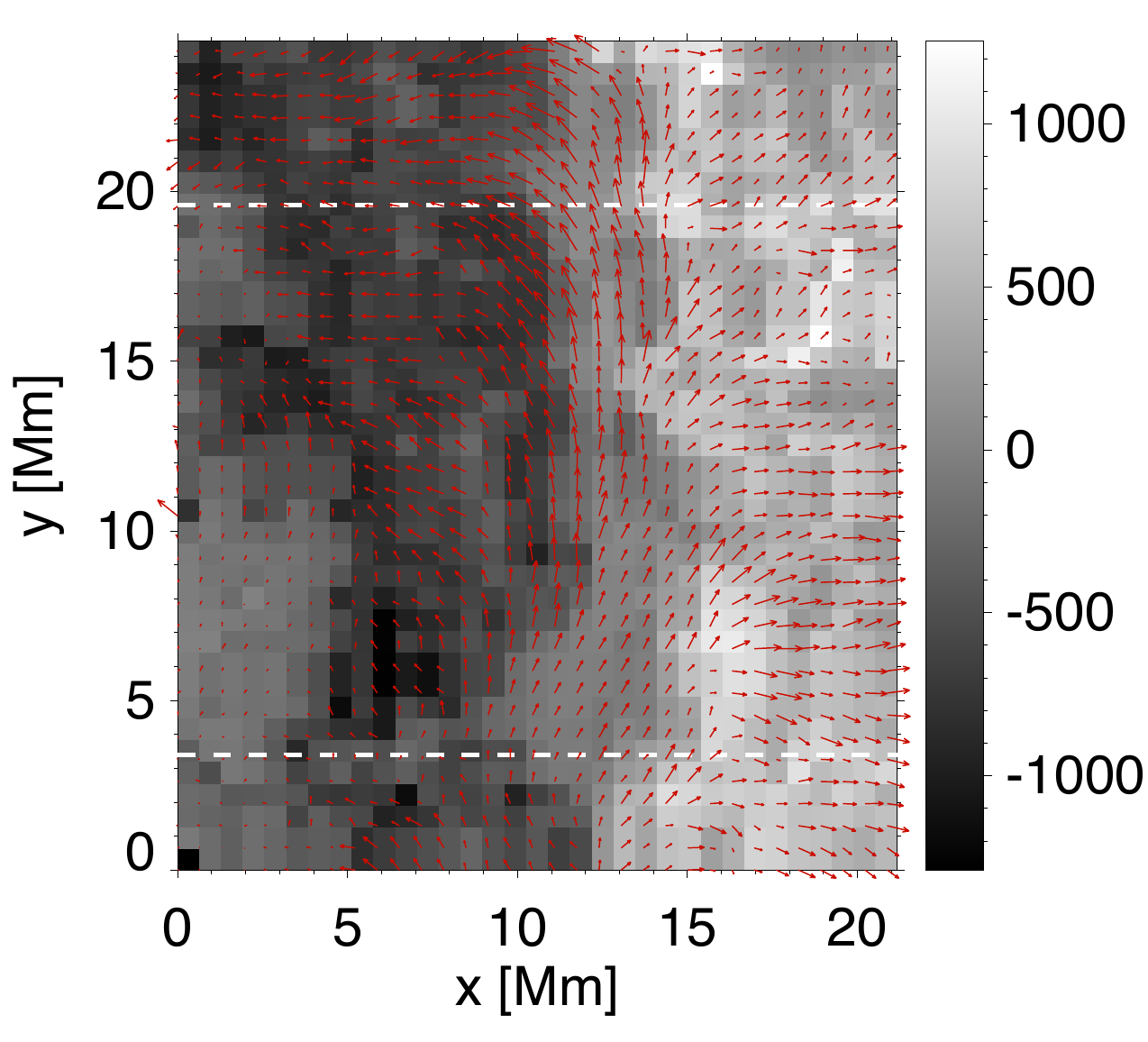}
\end{center}
\caption{Background : Vertical component of the magnetic field on July $5^{th}$ at the photosphere ($B_z ^{Si, O}$). The arrows represent the horizontal component of the magnetic field. The field strength is proportional to the
length of the arrows.} \label{fig2}
\end{figure*}

\begin{figure*} 
\begin{center}
\includegraphics[width=0.9\columnwidth]{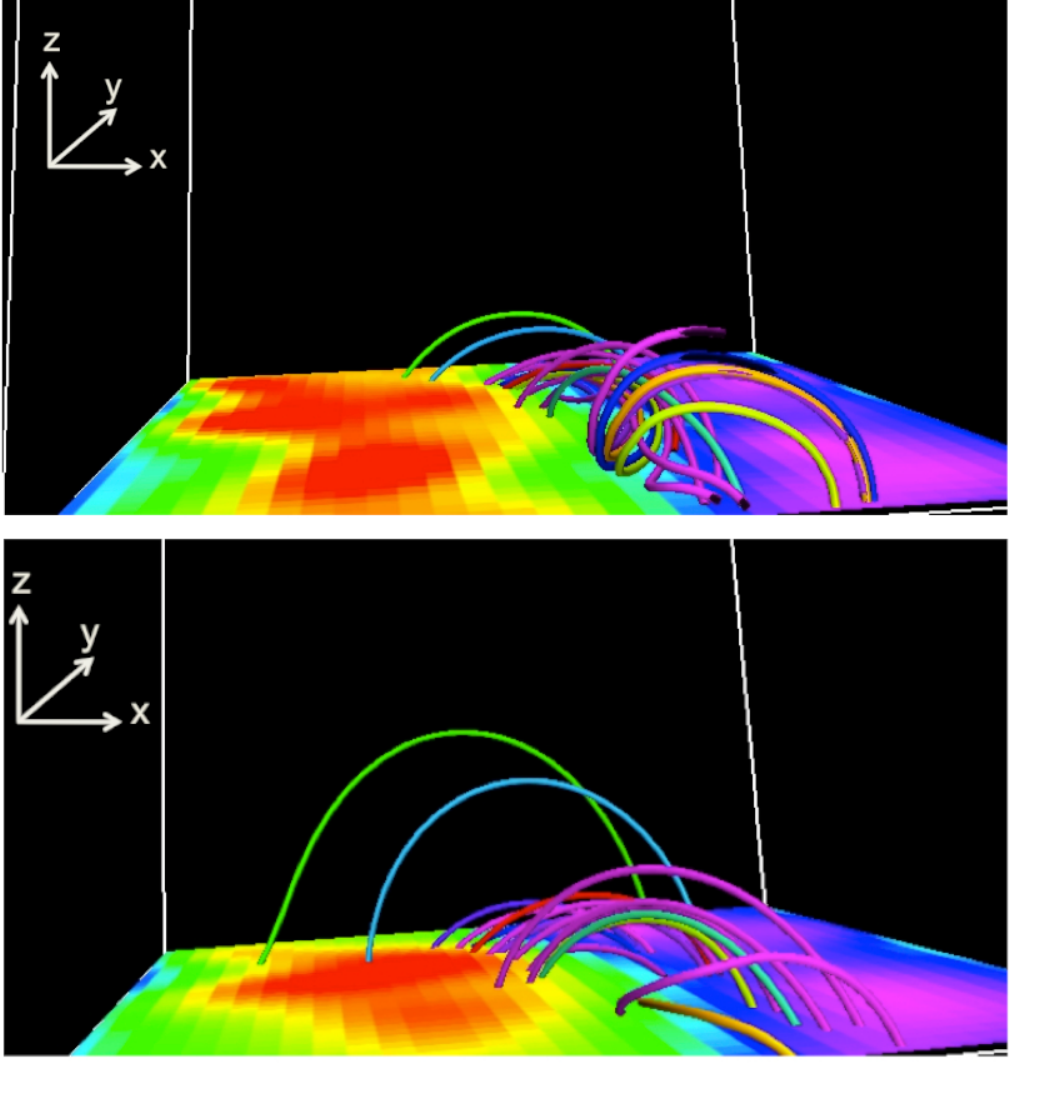}
\end{center}
\caption{Top: 3D view of the magnetic field of the filament on July $5^{th}$. The extrapolation
  results are computed using the observed photospheric magnetic field as lower
boundary condition. 
 Top background image:  vertical component of the magnetic field at the bottom of the extrapolation cube. Purple and blue: $B_z ^{Si, E}>0$;
  green and red:  $B_{z} ^{Si, E}<0$ 
 Bottom: Extrapolation results using the chromospheric magnetic field on July $5^{th}$ as lower
boundary condition. Bottom background image: the vertical component of the magnetic field at the bottom of the extrapolation cube $B_z ^{He, E}$ (same color scale as the top panel). The field lines are plotted in the filament region. The field lines that cross the lateral sides of the box are cut 2 pixels away from
  the boundaries to avoid plotting the part that might be affected by the lateral boundaries.} \label{fig3}
\end{figure*}

The vertical component of the magnetic field ($B_z ^{Si, O}$) inverted from the photospheric \ion{Si}{1} 10827 \AA\ data is plotted
in Figure~\ref{fig2} using a color scale. The lower case "z" in $B_z ^{Si, O}$ refers to the vertical component of the magnetic field,
while the upper case "O" and "Si" refer to the field obtained from Observations using the \ion{Si}{1} 10827 \AA\ line. The arrows in Fig.~\ref{fig2} indicate the direction of the horizontal magnetic field. The length of the arrows is 
proportional to the horizontal field strength. In the vicinity of the polarity inversion line, some arrows point from the negative polarity 
to the positive one, which is called inverse configuration. A {\it normal
  configuration} is obtained when the field arrows point from the
positive to the negative polarity.
The inverse configuration in the PIL region suggests that the filament above is formed 
by a twisted flux rope.

The extrapolated magnetic field is plotted in the top panel of Figure~\ref{fig3} using UCAR's
Vapor 3D visualization package (www.vapor.ucar.edu). This 3D view  shows a sample of magnetic field
lines in the filament region in order to outline its structure. The background map represents the
vertical component of the magnetic field ($B_z ^{Si, E}$) at the
bottom of the extrapolated cube.
The upper case "E" and "Si" refer to the field obtained by extrapolations using the \ion{Si}{1} 10827 \AA\ line vector magnetograms as lower boundary condition.
In the background map of the top panel of Figure~\ref{fig3}, purple and blue colors correspond to $B_z ^{Si, E}>0$; 
  green and red are reserved for $B_{z} ^{Si, E}<0$. The magnetic field forming the filament is distributed as a flux rope with dips in
  the lower half of the figure (i.e., with field line stretches which are
  convex as seen from the solar surface). This goes in the same direction as
  the findings of \citet{guo2010, canou10} and \citet{jing10}.

\citet{bard08}, have studied the properties of the \ion{Si}{1} 10827 \AA\ line in NLTE conditions.
They have determined using empirical models that the average formation height of \ion{Si}{1} 10827 \AA\
is about 320 km for a sunspot umbra and about 541 km in the quiet sun.     
In the case of the observed filament, we expect the average height of formation to be 
between these two boundary values. 

At this height of formation the photospheric plasma
does not have small values of the plasma $\beta$. Nevertheles, there are theoretical 
and observational indications \citep{wiegelmann2010a, martinez_gonzalez2010}
 that force-free extrapolations lead to a good retrieval of the actual magnetic field 
at higher atmospheric layers.


\subsection{Extrapolations starting from the chromosphere}

We have seen in the previous section, that the extrapolation studies have been conducted exclusively using photospheric spectral lines. 
Here we would like to present an extrapolation using the Helium triplet \ion{He}{1}  10830 \AA\  as lower boundary for the extrapolations.
Chromospheric lines as $\mathrm{H_\alpha}$ and \ion{He}{1}  10830 \AA\, have been used to identify filaments but never as boundary condition for extrapolations.

The bottom panel of Figure~\ref{fig3} shows a 3D display of a sample of field lines at the filament region. In this case the field lines harbor some shear 
in the lower half of the image but with no twist. Field lines have a normal configuration (from positive to negative), except at the very central part
of the filament (shown by the yellow field line) where the field lines are almost parallel to the filament axis. 
The bottom panel of Figure~\ref{fig3} suggests that the magnetic flux rope forming the filament (Lower panel of Fig.~\ref{fig3}) is located below the height of formation of the \ion{He}{1} 10830 \AA\ triplet. We will come back in detail to this point in the next sections.


\subsection{Height of coincidence between the photospheric and chromospheric magnetic fields} \label{sec:43}

\begin{figure} 
\begin{center}
\includegraphics[width=1.0\columnwidth]{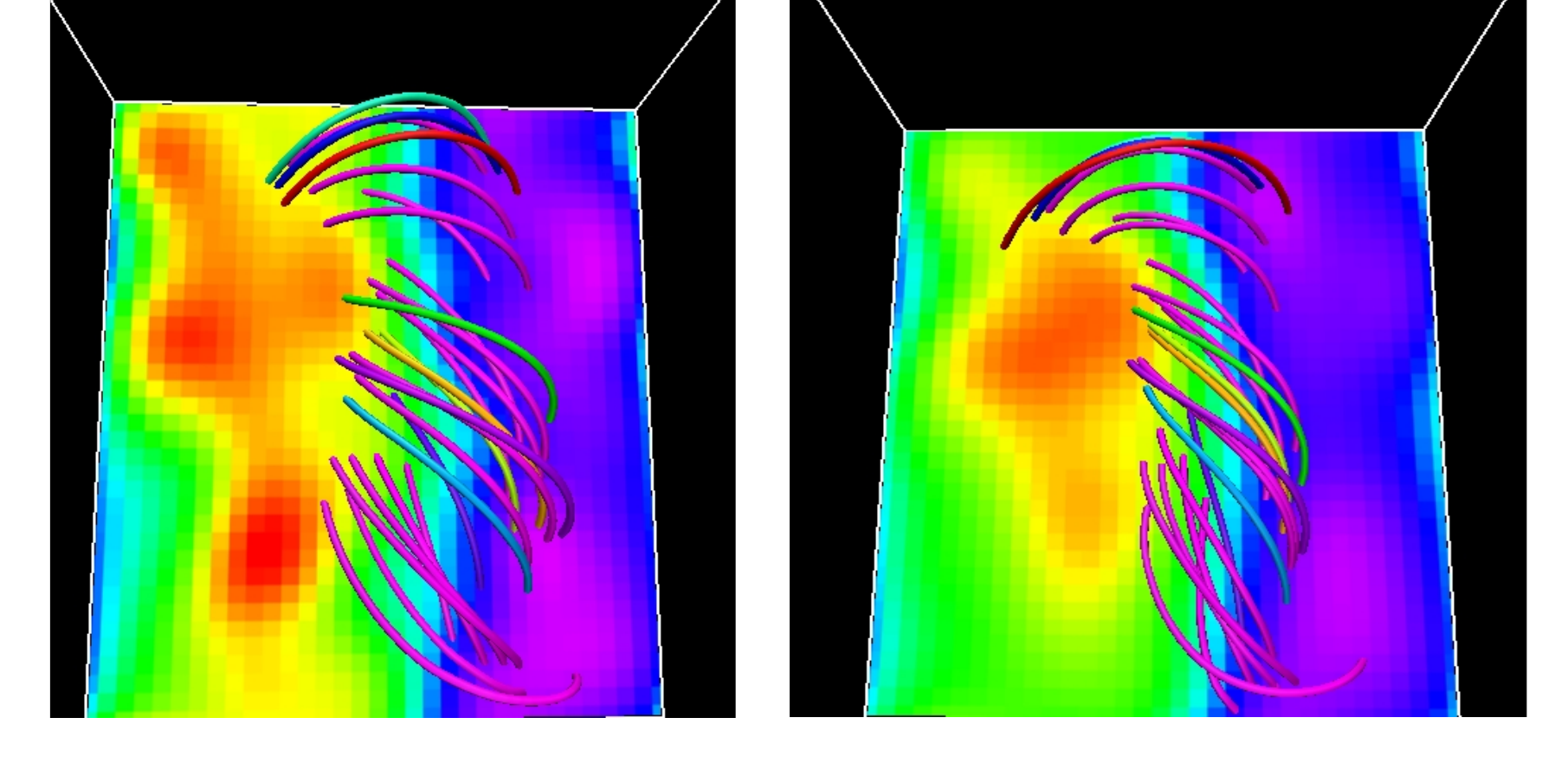}
\end{center}
\caption{Left panel: Field lines from the photospheric extrapolation. They are drawn above a minimum height of $\approx 1.6$ Mm above the photosphere (to make them comparable with the
chromospheric magnetic field). Right panel: Magnetic field lines from the
chromospheric extrapolations. They are drawn directly from the bottom of the corresponding extrapolation box (already at chromospheric height).
The two sets of field lines start inside a 2 Mm deep box placed along the filament. Field lines are allowed to end outside of the starting box, but are cut if they go below it. 
The background images are as described in Fig.~\ref{fig3},
and the data are from July $5^{th}$.} \label{fig5} 
\end{figure}

A natural question to ask when analyzing the extrapolations starting from the
photosphere and the chromosphere is: does the magnetic field from the two
extrapolations agree with each other?  In order to answer this question, we
first plot the field lines that result from the two 3D extrapolation boxes
(Fig.~\ref{fig5}). The field lines start inside a 2 Mm deep box which
extends along most of the filament. The field lines can develop outside of
the box but will be cut if they go below it.
This configuration allows
selecting field lines above a suitable height in order to make it possible to
compare field lines starting from the photosphere with those starting from the
chromosphere.

Therefore, the field lines in the left panel of Fig.~\ref{fig5} are all above an imaginary boundary situated at $\approx 1.6$ Mm above the bottom of the
extrapolation cube. We have tried various heights and found that this one ($\approx 1.6$ Mm) gives a good agreement between the field lines from the photospheric 
extrapolations (left panel) and the ones from the chromospheric extrapolations (right panel of Fig.~\ref{fig5}) which are drawn directly
from the lower chromospheric boundary. 
A visual inspection of both panels suggests a good agreement between the morphology of the magnetic field obtained from 
photospheric and chromospheric extrapolations.

\begin{figure} 
\begin{center}
\includegraphics[width=0.45\columnwidth]{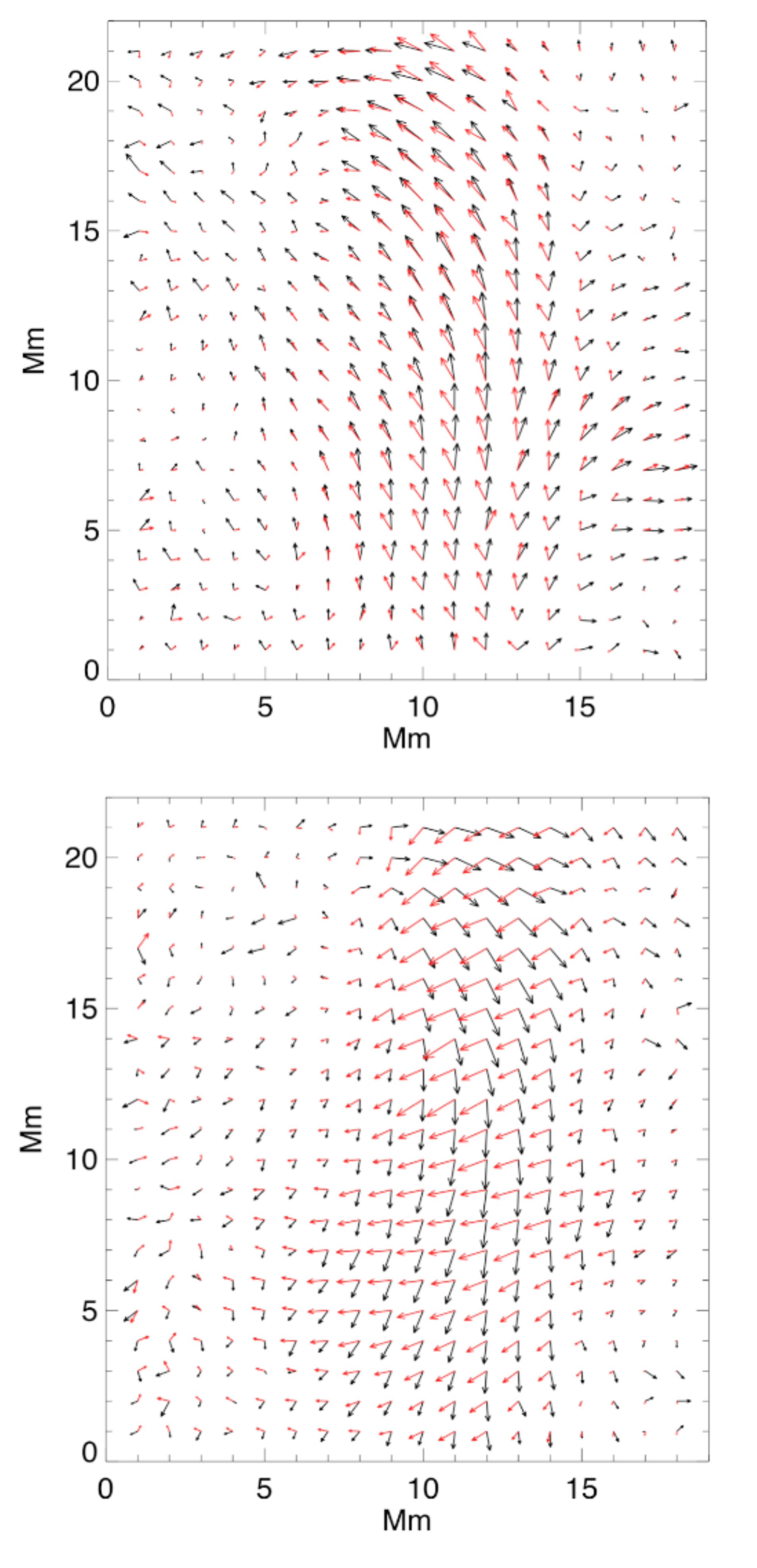}
\end{center}
\caption{Red arrows: Flow representation of the horizontal magnetic field from the photospheric extrapolations at the height where the extrapolated photospheric field strength matches the observed field strength in the Helium vector magnetogram. Black
arrows: Flow representation of the observed horizontal magnetic field in the \ion{He}{1}  10830 \AA\  vector magnetogram. Top panel: Solution obtained with the correct 180 degree ambiguity solution computed with the AZAM utility \citep[][and references therein]{lites05}. Note that the solutions are obtained for both photospheric and chromospheric fields. Bottom panel: Similar to the top panel but using the other solution of the 180 degree ambiguity. All data sets were taken on July $5^{th}$.} \label{fig7}
\end{figure}

To provide a quantitative comparison between the two sets of extrapolations, we compute the magnetic field strength in the photospheric extrapolation cube and look for the height
where this field strength matches the observed chromospheric one. This operation is repeated 
along each column of the photospheric extrapolation cube.  At the height ($\mathrm{H}$) where the two field strengths coincide, 
the magnetic field components $B_x ^{Si, C}$, $B_y ^{Si, C}$, $B_z ^{Si, C}$ are stored as 2D arrays (The uppercase "C" stands for Coincide).
For instance, the horizontal magnetic field from the photospheric extrapolation ($B_x ^{Si, C}$, $B_y ^{Si, C}$) across the field 
of view is plotted in Fig.~\ref{fig7} and represented by red arrows. The one from the observed Helium is plotted as black arrows. The arrows length is 
proportional to the horizontal field strength. The top panel of Fig.~\ref{fig7} shows a relatively good agreement between the two magnetic fields indicating that the observed
magnetic field $B^{He, O}$ is well matched by the extrapolated $B^{Si, E}$ field especially at the filament region. This goes in favor of validating the Silicon photospheric
extrapolations. The average height where $B^{Si, E}$ matches $B^{He, O}$ is $1.57$ Mm in the filament region, with a standard deviation of $0.66$ Mm.
Therefore, it is possible to determine an average formation height of the \ion{He}{1} 10830 \AA\ triplet. This would be $1.57 \,  \mathrm{Mm} \, +$ the formation height of the \ion{Si}{1} 10827 \AA\, which leads to 
approximatively $2$ Mm above the solar surface ($\tau = 1$).

\subsection{A complementary method for solving the 180 degrees ambiguity}

The top panel of Figure~\ref{fig7} indicates that the magnetic field extrapolated from the photosphere matches well
the one observed at the chromosphere. It is nevertheless interesting to redo the calculations
but for the case where the observed horizontal magnetic field is taken from the 
solution with 180 degrees shift of the azimuth in the LOS frame. This is done to test the coherence of the results with the other magnetic field
configuration.

From the lower panel of Figure~\ref{fig7} it is found that, in this case, the magnetic field extrapolated from the photosphere and the one
observed at the chromosphere have a quite different orientation in the
horizontal plane although they have 
similar field strengths at the height of coincidence. This clearly discards the solution with 180 degrees shift and provides a 
complementary method to test the validity of the chosen horizontal field as a result of the various routines 
used to solve the 180 degrees ambiguity of the horizontal magnetic field.


\subsection{Location of the dips along the filament field lines}

\begin{figure} 
\begin{center}
\includegraphics[width=1.0\columnwidth]{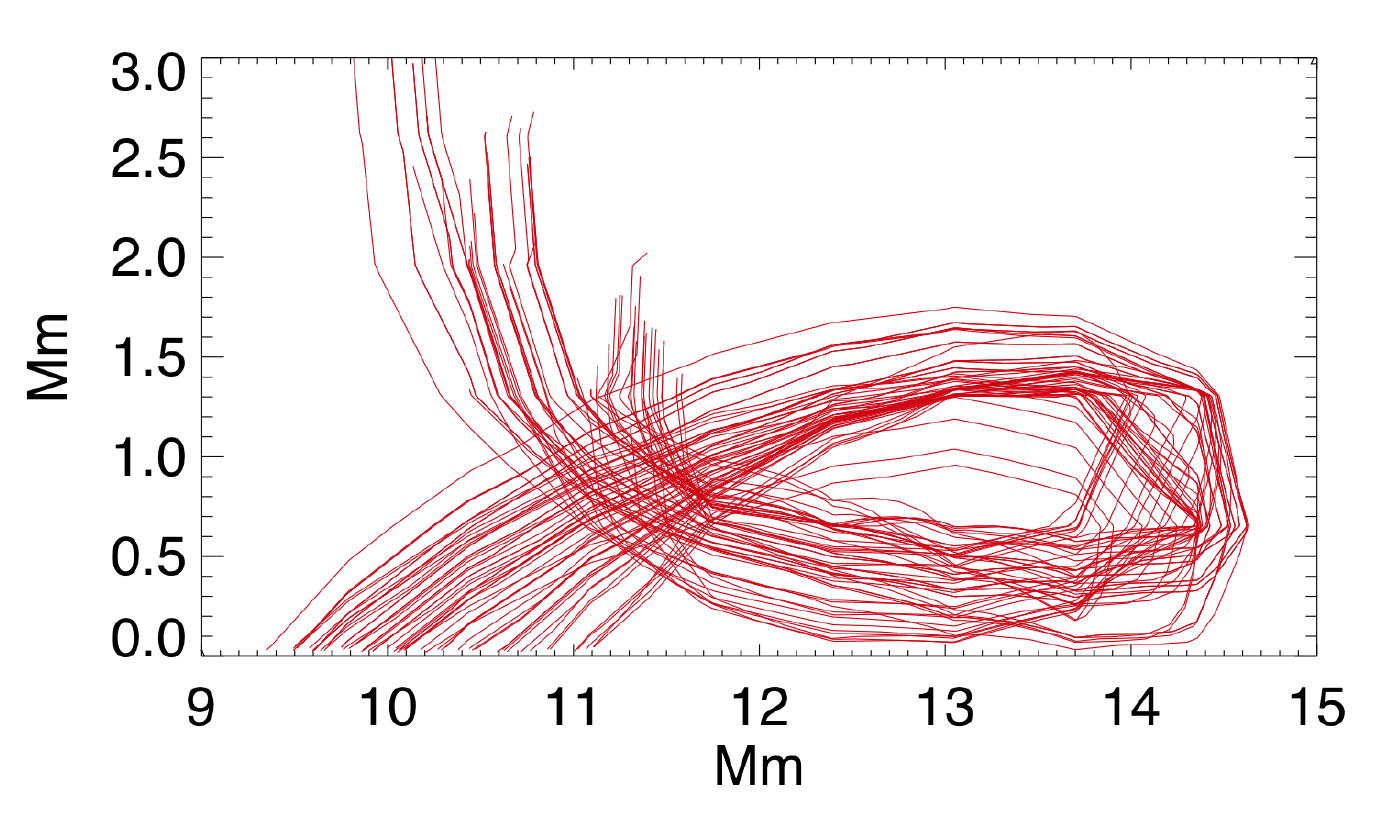}
\end{center}
\caption{X-Z view of magnetic field lines having a dip of at least 650 km depth. The computations use the 3D magnetic field extrapolated from the \ion{Si}{1} 10827 \AA\  vector magnetogram. This reveals the morphology of the flux rope with field lines making more than a full turn. 
In order to understand the morphology of this twisted flux rope, it is useful to have a look at the top panel of Fig.~\ref{fig3} showing another perspective of the filament magnetic field lines.} \label{fig9}
\end{figure}

The material carried by a filament must be sustained against gravity by the magnetic
curvature force associated with the dips in the field lines.
Using the extrapolated magnetic field from the Si vector magnetogram, we compute all the field lines that have a dip of at least 650 km depth (equivalent to the size of one grid cell). 
Figure~\ref{fig9} indicates that such field lines with dips exhibit a twisted structure, revealing the 
topology of the filament's magnetic field.  This X-Z view shows that the whole structure is lying 
relatively low. The axis of the filament is located some 1 Mm above the formation height of the 
Silicon line. 
This indicates that the axis of the TFR forming the filament is located below the formation 
height of the Helium triplet. 
The lower part of the field lines in Figure~\ref{fig9} exhibits an inverse configuration (from negative to positive)
whereas the upper part has a normal configuration. 
Recall that the average formation height of the Helium triplet (Sec.~\ref{sec:43}) in the filament region is $1.57$ Mm above the formation height 
of the Silicon line. At that height the field lines have a normal configuration. 
This helps understanding the lower panel of Fig.~\ref{fig3}, where the Helium
extrapolations exhibit no twist, which also fits with the presence of a normal configuration magnetic field at the lower boundary of the extrapolation box. 
It can be seen from Fig.~\ref{fig9} that the field lines make slightly more than one turn. These field lines extend over a horizontal distance of $\approx 18$ Mm. This leads to $\approx 0.055$ turns per Mm.


Following each field line in Fig~\ref{fig9}, it is possible to calculate the location of the lowest part of the dips.
These places are likely to be the preferred location where the plasma inside the filament would be gathered.
\citet{guo2010} and \citet{canou10} have found a correspondence between the location of dips
and the filament as observed in $\mathrm{H_{\alpha}}$. It is reasonable to argue that in order to build up opacity (e.g. in the
$\mathrm{H_{\alpha}}$ spectral region), filaments are filled with material which will be preferentially located in dips.


\begin{figure} 
\begin{center}
\includegraphics[width=1.0\columnwidth]{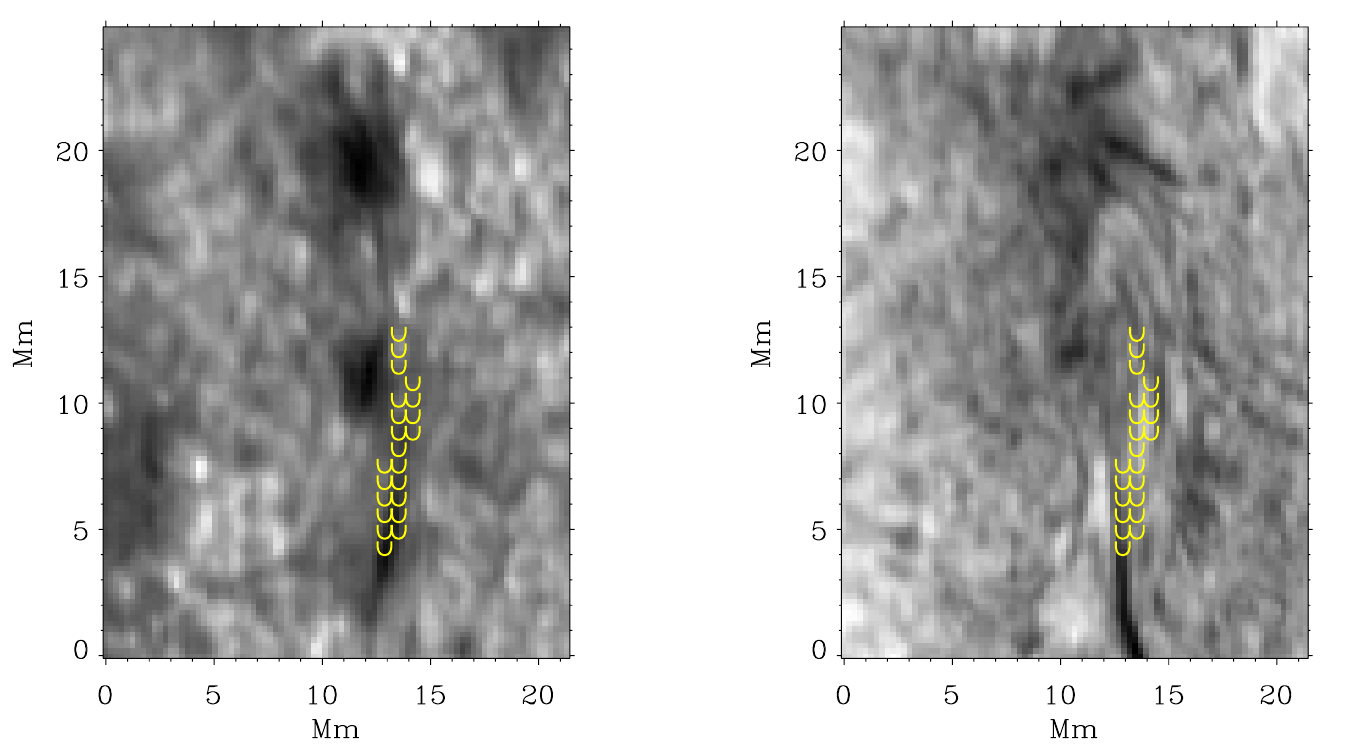}
\end{center}
\caption{Location of dips (drawn in yellow "U" shapes). These dips are calculated from the photospheric extrapolations for both panels. Left panel: locations of dips on top of \ion{Si}{1} 10827 \AA\ line core intensity.
Right panel: locations of dips on top of the red core of \ion{He}{1} 10830 \AA. The data on both panels are from the $5^{th}$ of July.} \label{fig11}
\end{figure}

\begin{figure} 
\begin{center}
\includegraphics[width=1.0\columnwidth]{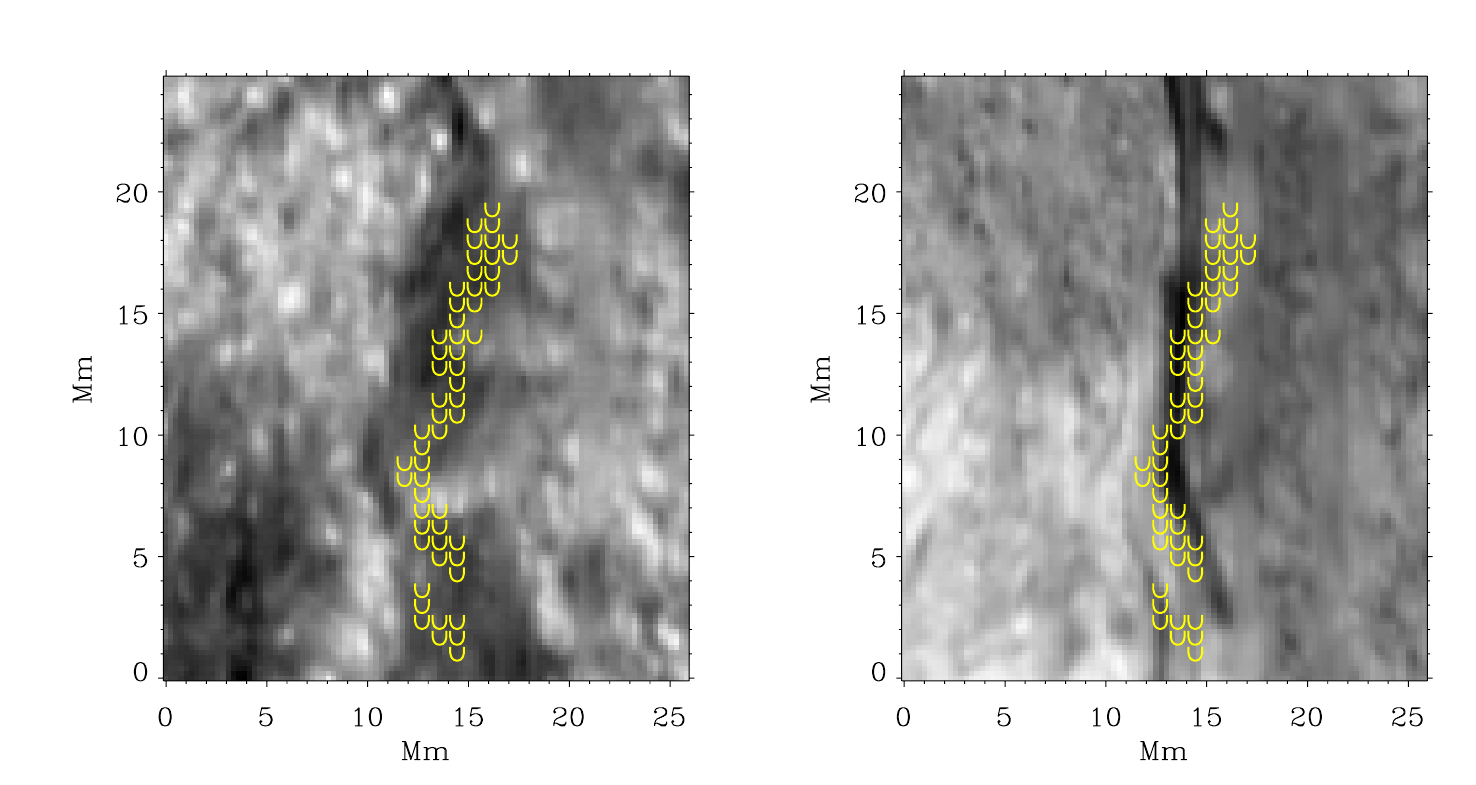}
\end{center}
\caption{Similar to Fig.~\ref{fig11} using a snapshot from the $3^{rd}$ of July (see Fig.~\ref{fig1} ).} \label{fig12}
\end{figure}

Figure~\ref{fig11} shows the locations of dips on top of the \ion{Si}{1} 10827 \AA\ line core intensity image (left panel) and the intensity 
image of the red core of the \ion{He}{1} 10830 \AA\ triplet (right panel). There is an agreement between the location of the filament and the dips, which 
are located in the lower half of the image where the filament has a flux rope structure.
There are no dips in the upper part of the panel since the magnetic field in this region has a normal configuration with no flux rope.  
An even  clearer case is seen in Fig.~\ref{fig12} where we repeated the same calculations as for Fig.~\ref{fig11} but for a
snapshot taken on the $3^{rd}$ of July. Note the correspondence with both
line cores. This analysis also provides a further indication on the location of the filament
(as the $\mathrm{H_{\alpha}}$ spectral signature of a filament), and allows also to corroborate the extrapolation results since the spectral signature 
of the line cores of \ion{Si}{1} 10827 \AA\ and \ion{He}{1} 10830 \AA\ are independent of the extrapolation analysis which involves the magnetic 
field and the force-free condition.         

\subsection{Testing the effect of the reduced field of view on the extrapolations} \label{sec:46}

\begin{figure} 
\begin{center}
\includegraphics[width=0.75\columnwidth]{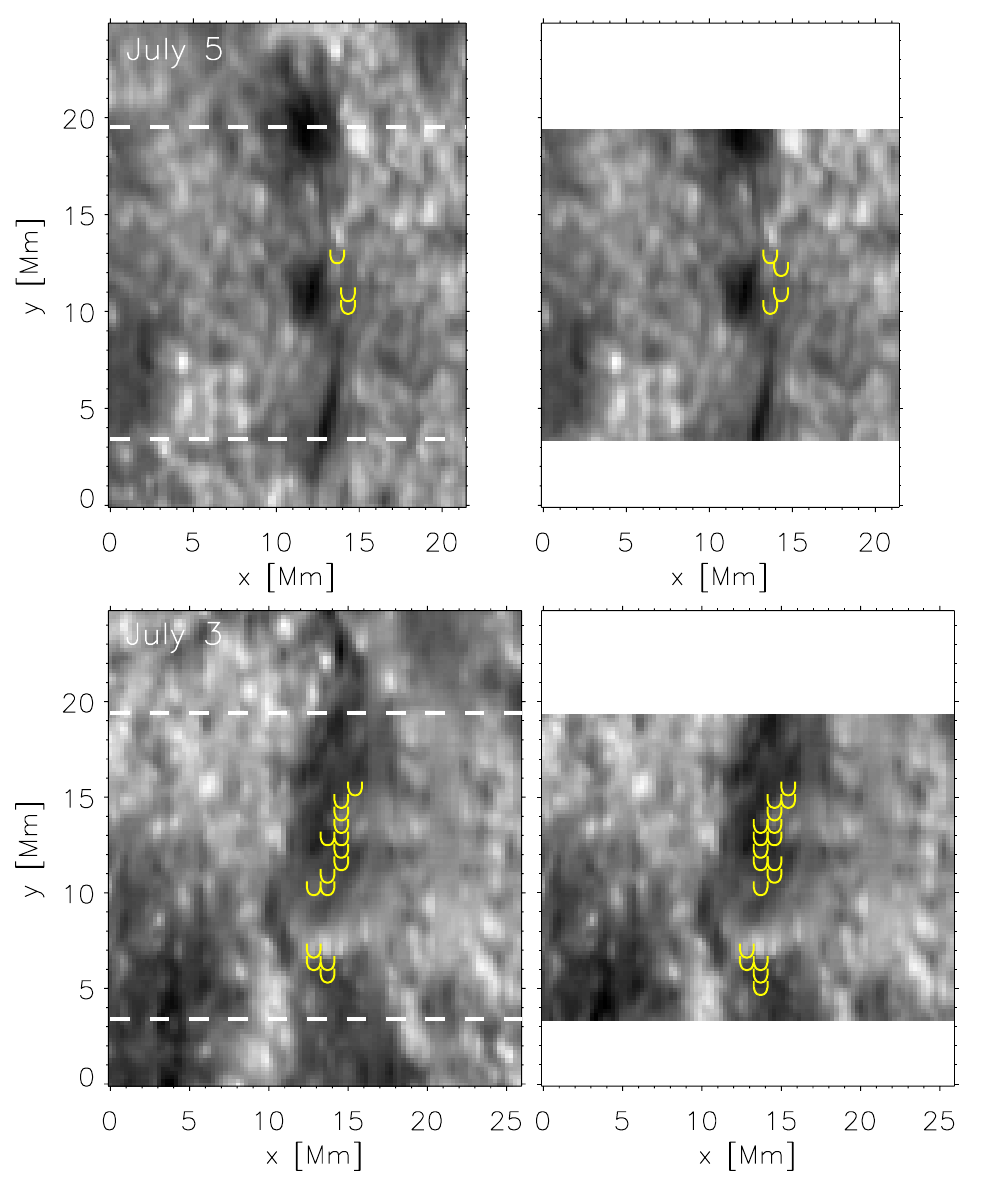}
\end{center}
\caption{Location of dips calculated from the photospheric extrapolations. 
Top row: locations of dips on top of \ion{Si}{1} 10827 \AA\ line core intensity on the $5^{th}$ of July.
Lower row: locations of dips on top of \ion{Si}{1} 10827 \AA\ line core intensity on the $3^{rd}$ of July.
Left column: the extrapolation of the whole field-of-view of \ion{Si}{1} 10827 \AA\ vector magnetogram is used, but the field line calculation to find the locations of dips start from the lower dashed line until the upper dashed line. This covers a portion similar to the reduced field-of-view test. 
Right column: the extrapolations of the reduced field-of-view of \ion{Si}{1} 10827 \AA\ vector magnetogram are used (same size as displayed in the background Silicon line core image). Dips are calculated in this reduced field-of-view.}  \label{fig104}
\end{figure}

\begin{figure} 
\begin{center}
\includegraphics[width=1.0\columnwidth]{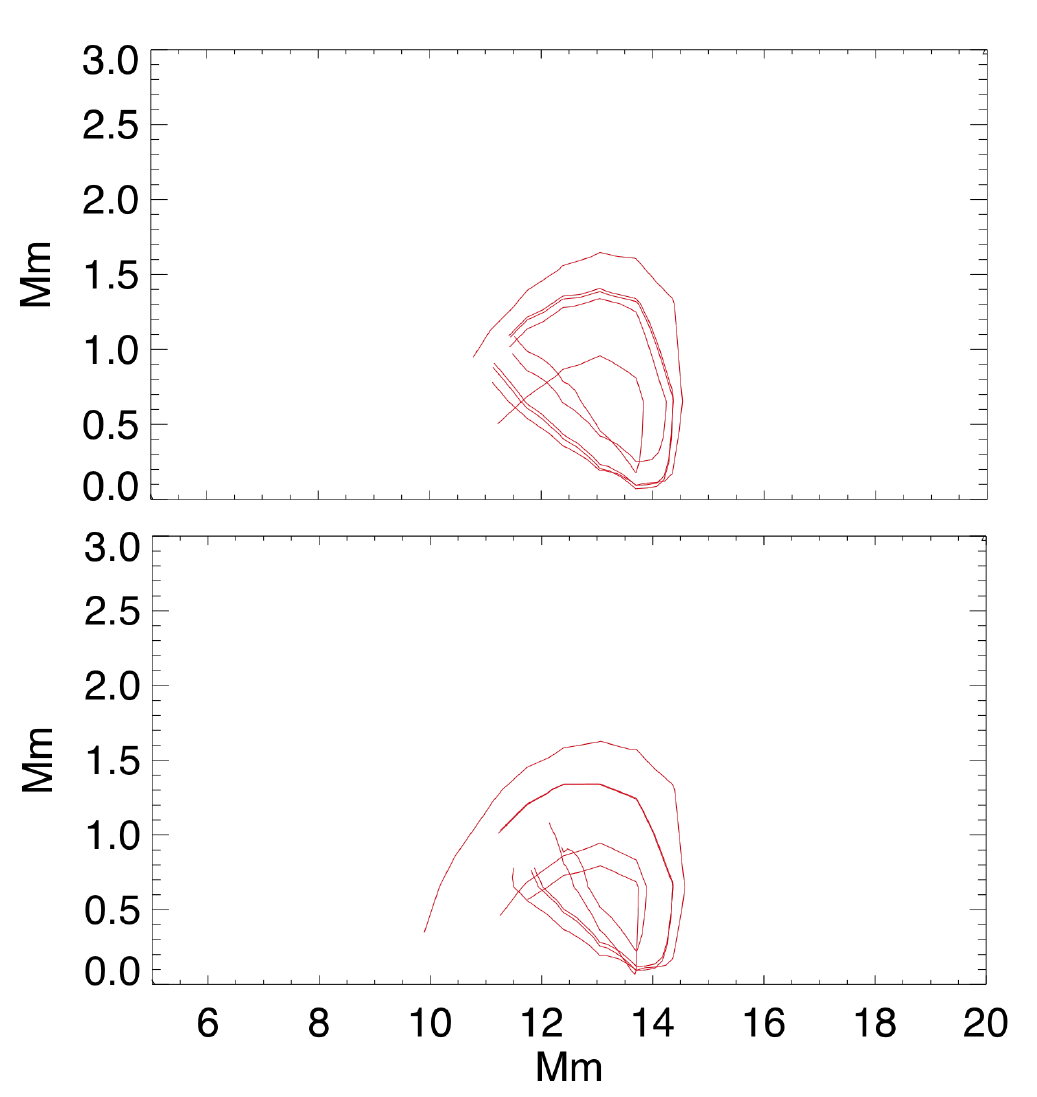}
\end{center}
\caption{Lower panel: field lines with dips calculated in the extrapolation with reduced field of view corresponding to the region indicated in the
upper-right panel of Fig.~\ref{fig104}. 
Upper panel: field lines forming dips computed from the extrapolation with the whole field of view, but calculated inside the region includes between the two dashed lines in the upper-left panel of Fig.~\ref{fig104}.} \label{fig14}
\end{figure}

The observations made with TIP-II have a somewhat reduced field of view.
Ideally the field of view should include the whole filament or even the whole active region (in general, a flux balanced region).
The fact that we use part of the filament might influence the large scale magnetic connectivity through
the filament. It is very useful to make quantitative tests to clarify the effects of reduced field-of-view \citep[private communication,][]{aulanier11}. 
We present here a test in which we cut part of the observed filament and
analyze how that affects the extrapolations. 
The chosen reduced field-of-view region is the one delimited by the two dashed lines in Fig.~\ref{fig2}. The resulting region is the one shown in 
the top-right panel of Fig.~\ref{fig104} corresponding to the $5^{th}$ of July. Similarly, the reduced field-of-view portion for the $3^{rd}$ of July is shown in the
lower-right panel of Fig.~\ref{fig104}. On these two panels we have drawn the location of dips using yellow "U" shapes. 
For the left column of panels in Fig.~\ref{fig104} we used the extrapolation of the whole field-of-view of \ion{Si}{1} 10827 \AA\ vector magnetogram, but the field line calculation to find the locations of dips start from the lower dashed line until the upper dashed line. This covers a region similar to the reduced field-of-view cases on both the $3^{rd}$ (bottom-left) and $5^{th}$ (upper-left) of July.
Figure~\ref{fig104} exhibit less locations of dips than the cases of Figs.~\ref{fig11} and~\ref{fig12}. This is primarily due to the fact that the field lines in the reduced field-of-view portion are shorter than the ones in the original full field-of-view case, and therefore there are less field lines that
fulfill the criteria of having dips of 650 km depth.   
A comparison between the left and right panels in Fig.~\ref{fig104} shows that they have some similarities indicating that decreasing the field of view alters only slightly the structure of the obtained filament and more precisely the location of dips. Figure~\ref{fig14} provides a closer look to the field lines forming the dips plotted in the upper panels of Fig.~\ref{fig104}. The field lines in the upper/lower panels of Fig.~\ref{fig14} correspond to the upper left/right panels of Fig.~\ref{fig104} respectively. 
The two sets of field lines
are very similar indicating that the fact of reducing the field of view does not severely affect the magnetic 
structure of the extrapolated filament. Nevertheless, a more complete test of the effect of field of view 
on the results of extrapolation of active region filament would be useful and interesting. That would include the
observation of a whole filament (ideally in an isolated and flux balanced region) and repeated extrapolations while reducing gradually the field of view. 
In the present work we were limited with the original field of view, therefore we could not run such a test. 
We intend to include a test of that sort as part of a future investigation. 
This test is complementary to the more complete series of tests presented in \citep[e.g.][]{schrijver06, schrijver08, metcalf08, derosa09}.

\subsection{Magnetic field increasing from the photosphere upwards}

\begin{figure} 
\begin{center}
\includegraphics[width=0.6\columnwidth]{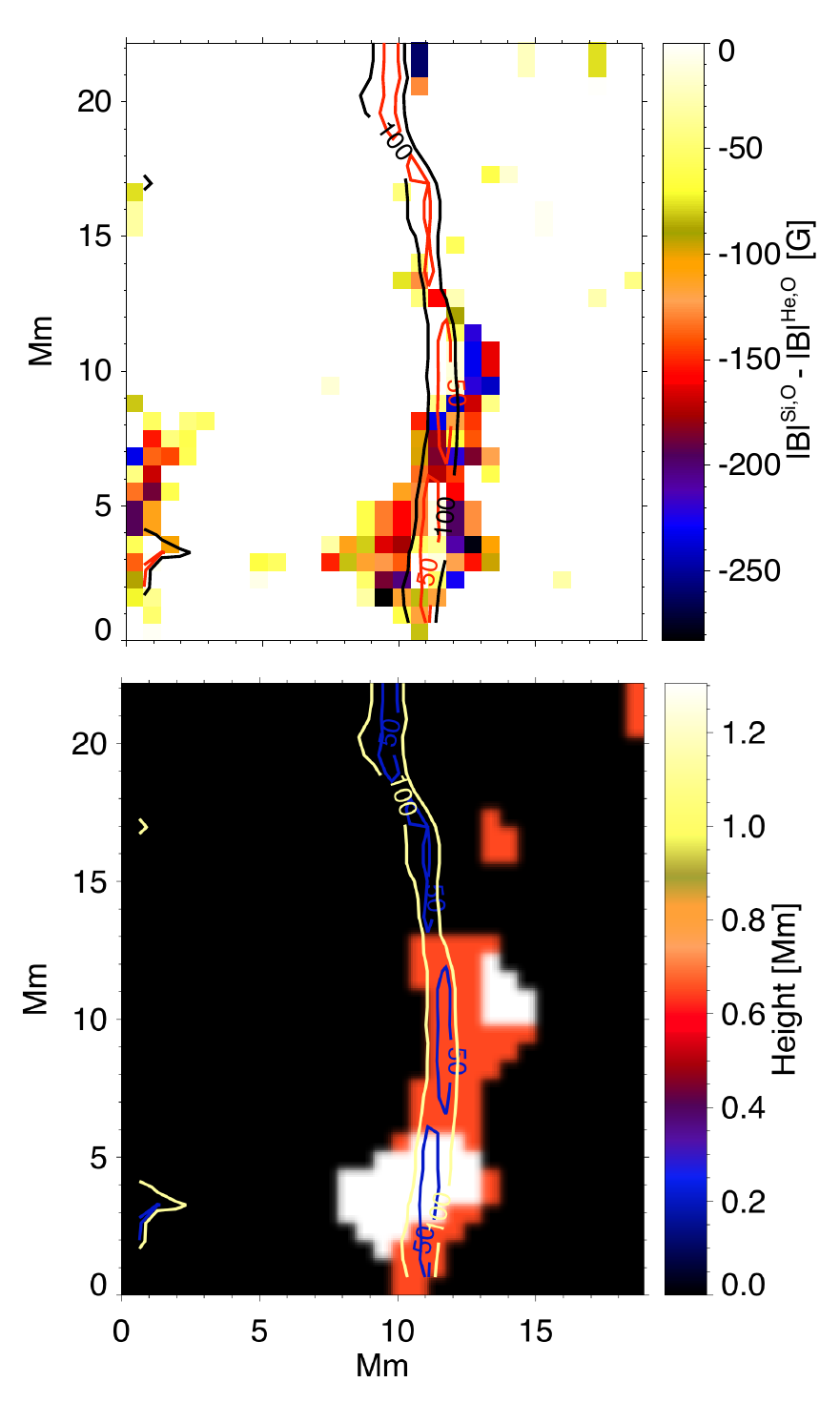}
\end{center}
\caption{Upper panel: Difference between the observed $|B|^{Si, O}$ and $|B|^{He, O}$. The positive values are clipped to $0$ in order to clearly see the negative values.
Lower panel: The height where the magnetic field strength from the photospheric extrapolations reaches a maximum. The data on both panels are from the $5^{th}$ of July.
The location of 
the neutral line is approximately indicated by the contour-plots of $B_{z}
^{He, O}$ shown on both panels.} \label{fig15}
\end{figure}

\begin{figure} 
\begin{center}
\includegraphics[width=1.0\columnwidth]{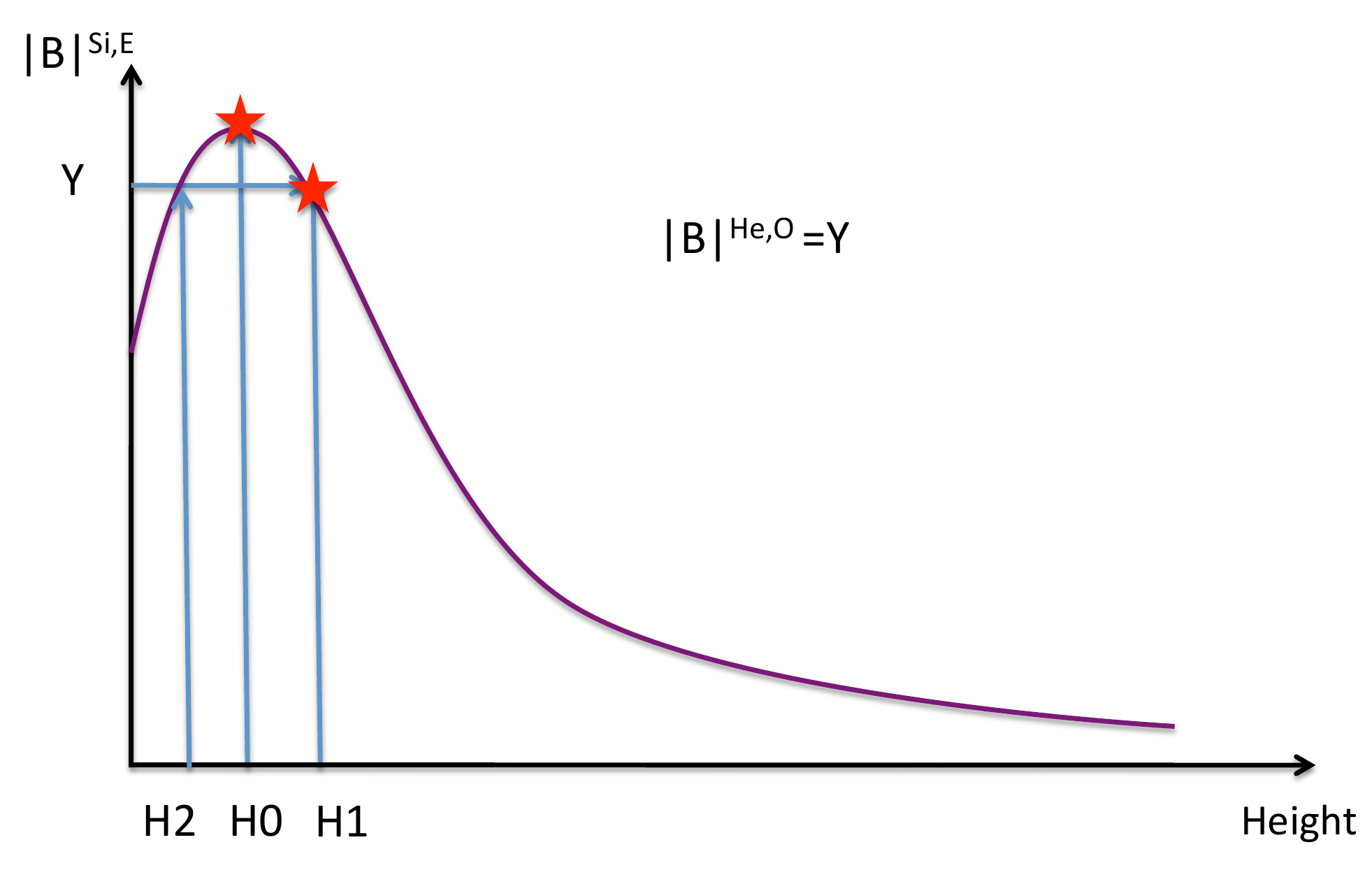}
\end{center}
\caption{Schematic explanation of the situation observed in the pixels where 
the extrapolated magnetic field increases upwards. First it increases reaching a maximum at the height "H0" plotted in the lower panel of Fig.~\ref{fig15}, then the magnetic field decreases again reaching the value
of the observed chromospheric magnetic field at the height "H1" } \label{fig16}
\end{figure}

Due to magnetohydrostatic equilibrium, the field of concentrated
vertical magnetic structures usually decreases with height above the  
photosphere. This is a typical case in a stratified atmosphere.
Nevertheless, vertical gradients of the field strength, 
indicating stronger fields in the upper layers, have been observed \citep[e.g., in polar crown prominences 
as shown by][]{leroy83} and modelled in the past \citep{aulanier03} and have been 
ascribed to the presence of dips \citep{anzer69, demoulin89}.
Here we would like to analyze the possibility of an upward increase of the
magnetic field strength in the filament region. 
The upper panel of Figure~\ref{fig15} shows the difference between the
observed field strengths of $|B|^{Si, O}$ and $|B|^{He, O}$. For the sake of
clarity, the figure only
shows negative values of those quantities.  A proxy of the location of 
the neutral line is indicated by the contour plots of $B_{z}
^{He, O}$ (i.e. small values of  $B_{z}^{He, O}$ coincide with the neutral line location). In the central part of the figure, the values 
of $|B|^{He, O}$ are larger than $|B|^{Si, O}$, indicating that the magnetic field in the chromosphere is larger than
the photospheric one, suggesting that the magnetic field has increased upwards. In order to understand this
phenomenon we look for the height where the extrapolated magnetic field from the photosphere reaches its maximum along each column.
This height is plotted in the lower panel of Figure~\ref{fig15}. It can be seen in the central region of the figure that the magnetic field reaches a maximum well above
the photosphere, at an average height of about $1$ Mm above the formation height of the \ion{Si}{1} 10827 \AA\ line. 

%
Figure~\ref{fig16} schematically explains the situation observed in those pixels where 
the magnetic field increases upwards. First the magnetic field increases up to a maximum reached at the height H0
(plotted, in fact, in the lower panel of Figure~\ref{fig15}), then the magnetic field decreases gradually until reaching the height of formation of the
Helium signal (H1) where the magnetic field is still larger than the one at the photosphere.
It is also seen that there are two solutions for the height of formation of the Helium signal (H1 and H2). The solution H2 is ruled out because it leads to abnormally low formation height. 
The H1 solution is kept and is consistent with the average height of formation of the Helium signal (see Sec.~\ref{sec:43}).

The reason why the magnetic field increases lies in the fact that the
corresponding part of the filament resembles the structure of a
force-free flux rope. In an ideal force-free flux rope, the azimuthal
component of the field yields an inward-pointing magnetic tension force
which is balanced by an increase of the magnetic pressure toward the rope
axis. This leads to an increase of the magnetic field strength in the central
regions of the rope.

\subsection{Complementary comments on the filament observed on the $3^{rd}$ of July}

Along the previous sections we have used observations taken on the $3^{rd}$ and $5^{th}$ of July, but with some focus on the filament as observed on the $5^{th}$ of July.
This has been done in order to convey a clear message and not overwhelm the reader with information about the time evolution of the filament (This time evolution will be discussed in \cite{kuckein11}). Additionally, the filament on the $5^{th}$ of July is very low lying (the filament's axis is located below the formation height of \ion{He}{1} 10830 \AA). This represents a peculiar and interesting case which we have chosen to address.
In order to complete the picture of the actually observed filament, we will briefly describe some further features of the filament as observed on the 3rd of July.

\begin{figure} 
\begin{center}
\includegraphics[width=0.9\columnwidth]{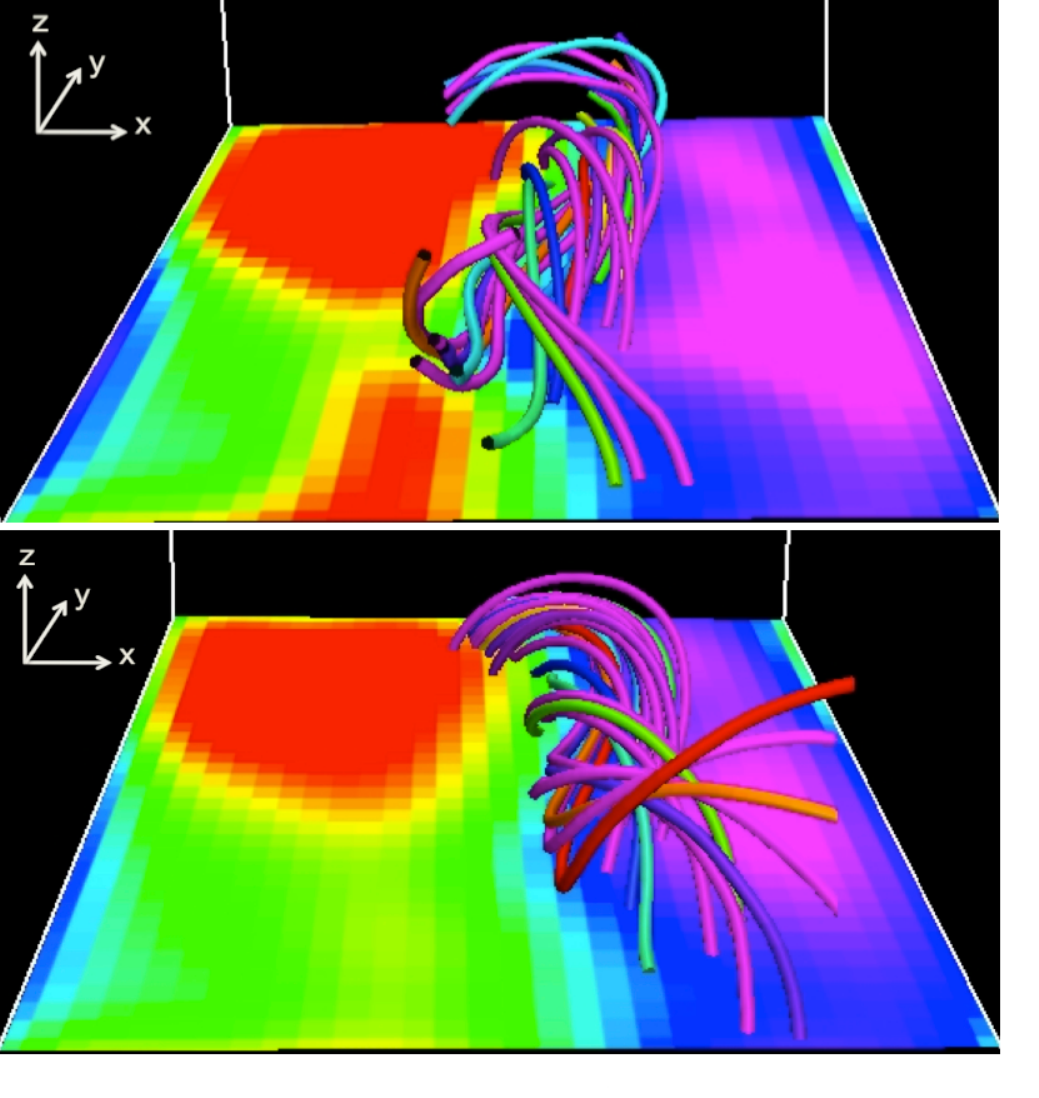}
\end{center}
\caption{Top: 3D view of the  magnetic field of the filament on July the $3^{rd}$. The extrapolation results use the photospheric magnetic field as lower
boundary condition. The sample of field lines is plotted in the filament region to outline its structure. Top background
image: vertical component of the magnetic field at the bottom of the extrapolation cube. Purple and blue: $B_z ^{Si, E}>0$;
  green and red:  $B_{z} ^{Si, E}<0$.
Bottom: Extrapolation results for the $3^{rd}$ of July using the chromospheric magnetic field as lower
boundary condition. Bottom background image: vertical component of the magnetic field at the bottom of the extrapolation cube $B_z ^{He, E}$
(same color scale as in the top panel). } \label{fig17}
\end{figure}

It can be seen in Fig.~\ref{fig12}, that the line core of the \ion{Si}{1} 10827 \AA\ and the red core of the \ion{He}{1} 10830 \AA\  show a spectral signature
reminiscent of the axis of the filament on July the $3^{rd}$. The locations of dips along the filament coincide well with the line core signature (Fig.~\ref{fig12}). The filament in this case is located
higher in the atmosphere and therefore imprints a clear spectral signature on both photospheric and chromospheric lines. 
The observed magnetic field from photospheric vector magnetograms
exhibits an inverse configuration (field lines going from negative to positive polarities) in the filament region. 
The observed chromospheric vector magnetograms indicate that the magnetic field in the filament region has a slightly inverse configuration with most of the field being parallel to the filament's axis.      
From extrapolations (Fig.~\ref{fig17}), it can be seen that the filament is formed
by a twisted flux rope. This is revealed from both photospheric and chromospheric extrapolations.

\section{Discussion and summary}

This paper is dedicated to building diagnostics about an active region (AR) filament using non-linear force-free extrapolations
starting from the photosphere and the chromosphere.  
Extrapolations of AR filaments are usually done using the photospheric magnetic field as a lower boundary condition. This is generally convenient
since the photospheric magnetic field is more accessible to observational techniques for various reasons, e.g. i) it has a large magnetic field 
implying a large Zeeman splitting, ii) the photon flux coming from the photosphere is large producing a high signal to noise ratio which makes it easily
detectable, iii) there are many available instruments measuring photospheric field, etc. It is nevertheless interesting to probe the properties of
filaments from two layers (or multi-layer if possible) since this approach provides several advantages like testing the validity
of the extrapolation solutions. It allows as well testing 
the consistency between the extrapolations starting from the photosphere and the chromosphere. We just began to see the potential of multi-layer extrapolations, e.g. the
determination of the matching height between the extrapolated photospheric field and the observed chromospheric one.
This can be used as a method to determine the relative height of formation of higher-lying lines/multiplets compared to lower-lying ones. 
We find that the average formation height of the \ion{He}{1} 10830 \AA\ triplet is about $2$ Mm above the surface of the sun. 
Multi-layer extrapolations provides also a complementary method for testing the $180^{\circ}$  ambiguity solutions. 

The AR filament studied here, harbors a TFR structure with a rather low-lying axis on the $5^{th}$ of July, where the axis is located at about $1.4$ Mm above the solar surface. This height is 
below the formation height of the Helium signal. This result is consistent with the fact that the magnetic field as observed in Helium magnetograms harbors a normal configuration. 

In the present study we have essentially focused on a snapshot taken on the 5th of July (see Figs. 1 and 2). We have also briefly discussed about a snapshot observed on the 3rd of July (Figs. 1 and 12). Recall that on the $3^{rd}$ of July, the filament's axis was higher located and clearly seen in the spectral signature of  \ion{Si}{1} 10827 \AA\ and \ion{He}{1} 10830 \AA\ line cores.  Actually, as discussed in section 2 (observations and data analysis), the observations used here belong to a time series of snapshots covering the $3^{rd}$ and $5^{th}$ of July 2005. In order to build a clear and detailed diagnostic we have essentially focussed on studying one case (on the 5th of July). A more detailed analysis about the temporal evolution of the filament will be addressed at a later stage \citep[see also][]{kuckein11}. Nevertheless, it is worth commenting on the other snapshots. An essential thing to mention, is that snapshots observed on the same day exhibit similar properties (height of location of the filamentÕs axis, magnetic field strength, etc). This means that the rest of the snapshots observed on the $5^{th}$ of July, have quite similar properties to the one described here, and similarly for the snapshots taken on the $3^{rd}$ of July.

For the filament on the $5^{th}$ of July, we have seen that, if taken separately, the photospheric and chromospheric extrapolations would lead to conclude that on the one hand the filament is formed by a TFR (from photospheric extrapolations), whereas on the other hand there is no TFR (from the chromospheric extrapolations). The multi-layer extrapolations helped solving this apparent discrepancy by showing that depending on the location of the TFR it might be seen at one layer or the other while the two extrapolations remain coherent with each other. Let us assume hypothetically that an active region filament is such that the lower part of the TFR is located at the upper chromosphere and does not extend to the photosphere. In this case, it would be seen in the chromospheric extrapolations, but the photospheric ones would not necessarily reveal it, especially since the photosphere is a relatively high plasma-$\beta$ medium and therefore the surrounding magnetic field in the AR might considerably alter the magnetic field near the PIL making the horizontal component of the magnetic field harbor a normal configuration. If we would only observe the photospheric field, this would lead to finding no TFR in the filament region, whereas in fact the filament's structure is a TFR. In fact this is true only if the electric currents present in the region are not reminiscent of a TFR, since it has been shown by \citet{valori10} that NLFFF extrapolations
have the potential to find a TFR even if its dips are not present at the magnetogram level. Nevertheless, this hypothetical scenario highlights the usefulness of multi-layer extrapolations to help probing the properties of filaments and extracting further information about the observed magnetic structure.             

In the case studied here, the material in the dense part of the filament, along its axis, is confined inside a flux rope where the field lines are bent
producing an excess of magnetic tension towards the center of the TFR. This tension is compensated by the outward oriented magnetic pressure
producing an enhancement of the magnetic field inside the TFR. This is reflected by a vertical positive gradient of the magnetic field from the photosphere 
up to about 1 Mm above the formation height of the \ion{Si}{1} 10827 \AA\ signal. In brief,
the field increases from the photosphere upward in the flux rope region.

There are only few studies dealing with extrapolations of AR filaments, the most recent being from \citet{guo2010, canou10} and \citet{jing10}.
These studies picture AR filaments
as TFR with the exception of \citet{guo2010} who have shown that TFR and dipped sheared arcades can co-exist in the same filament. We also find a flux rope from photospheric extrapolations, but have seen that the chromospheric extrapolations might show a different picture if the
observed spectral line/multiplet is formed above the filament's axis. There is a clear necessity of further observational campaigns with the appropriate extrapolations to further clarify the properties of AR filaments and their temporal evolution. Multi-layer extrapolations would be helpful to compare the resulting 3D structure of filaments with the variety of theoretical models existing in the literature \citep{mackay2010}.

\vskip 5mm

\acknowledgements{
Financial support by the European Commission through the SOLAIRE Network (MTRN-CT-2006-035484) and by the Spanish Ministry of Research and
Innovation through projects AYA2007-66502, AYA2007-63881, CSD2007-00050, AYA2009-14105-C06-03 and AYA2011-24808 is gratefully acknowledged.}

\end{document}